\documentclass[twocolumn,amsmath,amssymb,tighten,floatfix,prl]{revtex4-1}
\usepackage{graphicx}% Include figure files
\usepackage{dcolumn} % Align table columns on decimal point
\usepackage{bm, bbm}      % bold math
\usepackage{curves}
\usepackage{epic}
\usepackage{amssymb}
\usepackage{wasysym}
\usepackage{epsf, times}
\usepackage{subfigure}
\usepackage{eufrak}
\usepackage{amsthm}
\usepackage{mathrsfs}

\begin{document}

%\preprint{APS/123-QED}
\title{Optimal control for unitary preparation of many-body states: application to Luttinger liquids}

\author{
Armin Rahmani
and
Claudio Chamon
       }
\affiliation{
Physics Department, Boston University, Boston, MA 02215, USA
            }

\date{\today}

\begin{abstract}
Many-body ground states can be prepared via unitary evolution in cold atomic systems. Given the initial state and a fixed time for the evolution, how close can we get to a desired ground state if we can tune the Hamiltonian in time? Here we study this optimal control problem focusing on Luttinger liquids with tunable interactions. We show that the optimal protocol can be obtained by simulated annealing. We find that the optimal interaction strength of the Luttinger liquid can have a nonmonotonic time dependence. Moreover, the system exhibits a marked transition when the ratio $\tau/L$ of the preparation time to the system size exceeds a critical value. In this regime, the optimal protocols can prepare the states with almost perfect accuracy. The optimal protocols are robust against dynamical noise. 
\end{abstract}

\maketitle
%
%
%%%%%%%%%%%%%%%%%%%%%%%%%%%%%%%%%%%%%%%%%%%%%%%%%%%%%%%%%%%%%%%%%%%%%%%%%%%%%%

Loading cold atomic gases into optical lattices provides an opportunity to study the nonequilibrium properties of quantum matter in thermally isolated and highly tunable environments~\cite{Greiner2002,*Kinoshita2006,*Sadler2006,*Hofferberth2007,Bloch2008}. 
The central object of such studies is a many-body quantum state which undergoes unitary evolution generated by a time-dependent local Hamiltonian.

 In an important class of problems, we are interested in using the unitary evolution to transform an initial state that is the ground state of a local Hamiltonian to the ground state of another Hamiltonian. Such problems are relevant for the preparation of states in regimes where direct equilibration is difficult~\cite{Duan2003,Garcia-Ripoll2004,Sorensen2010}. Ground states preparation is the key to simulating many-body model Hamiltonians, such as the Hubbard model, which can be realized with cold atoms~\cite{Jaksch2005,Lewenstein2007}.

If one had infinite time to wait, according to the adiabatic theorem of quantum mechanics, the unitary transformation can be done with arbitrary accuracy in any finite system. Extrinsic losses and quantum decoherence, however, set an upper bound on the practical time to carry out the process. In any finite time $\tau$, nonadiabatic effects are unavoidable~\cite{Polkovnikov2010}. These effects are most severe in the absence of an energy gap~\cite{Polkovnikov2008}. In this paper we focus precisely on dynamics within the gapless phase by studying Luttinger liquids.

Let us start by casting the question of finding the optimal dynamical protocol in a generic way. Assume we have a local Hamiltonian $H(\{g\})=\sum_{i=1}^M\:g_i\: \hat{O}_i$
where the $\hat{O}_i$s are local operators and the $g_i$s are coupling constants that, within a given range, can be tuned to any value as a function of time. We would like to transform $|\Psi_1\rangle$ which is the ground state of $H(\{g_1\})$ to $|\Psi_2\rangle$, the ground state of $H(\{g_2\})$ in a given time $\tau$. How close can the final state
$
|\Psi(\tau)\rangle={\cal T}\exp[{-i\int_0^\tau dt^\prime H(\{g(t^\prime)\}) }]|\Psi_1\rangle
$
be to the desired ground state $|\Psi_2\rangle$? Here ${\cal T}$ represents time-ordering.

 The meaning of closeness above depends on the measure used. There are several popular measures like the excess energy for example~\cite{Eckstein2010}. Here we use the wave function overlap
\begin{equation}\label{eq:overlap}
{\cal F}[\{g(t)\}]=|\langle\Psi(\tau)|\Psi_2\rangle|^2.
\end{equation}
The problem is then reduced to finding the time-dependent $\{g(t)\}$ that maximizes the functional above. This interesting question in quantum dynamics~\cite{Polkovnikov2010} is in fact a typical problem in optimal control theory as noted in Ref.~\cite{Caneva2009,Doria2010}. Let us emphasize that we are concerned only with the final state and maintaining adiabaticity during the evolution, as for example in Ref.~\cite{Rezakhani2009}, is not a constraint. Moreover, we are restricted to \textit{local} Hamiltonians allowed by the experimental setup.  

The focus of this work is the Luttinger model for interacting fermions in one space dimension, whose Hamiltonian can be written in momentum space as follows
\begin{equation}\label{eq:H-momentum-space}
H=u\sum_{q>0} \left(K \:\Pi_q \Pi_{-q}+ \frac{1}{K}\:q^2\:\Phi_q \Phi_{-q} \right)
\end{equation}
where $\Phi_q$ are bosonic fields and $\Pi_q$ their conjugate momenta. 
%with $ \left[\:\Phi_q,\Pi_{q^\prime} \:\right]=i \delta_{q q^\prime}$
The parameters $u$ and $K$ are respectively the velocity of the charge carriers and the Luttinger parameter. 
We consider a fixed number of particles, focusing on the half-filled case. The zero mode, which is responsible for changing the particle number sector in the bosonized description, is therefore excluded in the above expression. 
Assuming we have an odd number of sites $L$ in the system, the momenta $q$ are given by $q=2\pi\frac{n}{L}$ for $n=1 \cdots, \frac{L-1}{2}$.

This Luttinger Hamiltonian is the low-energy effective theory for models of interacting fermions on a one-dimensional lattice, in particular the 1D Hubbard model
\begin{equation}\label{eq:Hamiltonian}
H=\sum_{j}\left[- c^{\dagger}_{j} c_{j+1}+{\rm h.c.}+V (n_{j}-\frac{1}{2}) (n_{j+1}-\frac{1}{2}) \right].
\end{equation}
Since spin degrees of freedom are not essential for our discussion, here we focus on spinless fermions. With the hopping amplitude set to unity as in Eq.~(\ref{eq:Hamiltonian}), the Luttinger parameters $u$ and $K$ are related to $V$ via the Bethe ansatz (see Ref.~\cite{Rahmani2010} for example).

 We will consider trajectories for $u$ and $K$ that are parametrized by a time-dependent $V(t)$: $u=u(V)$ and $K=K(V)$. The strength of Hubbard-type interactions can be tuned in optical lattices both by manipulating the optical potential with lasers~\cite{Olshanii1998,Jaksch1998,Giorgini2008} and through Feshbach resonances controlled with magnetic fields~\cite{Jordens2008,Giorgini2008}.

 Notice that while the relation between the Luttinger liquid (LL) and the 1D Hubbard model holds at low energies, the results we find for the optimum dynamical protocol in the former should be applicable to the latter when the total momentum $q n_q$ in each harmonic oscillator mode in Eq.~(\ref{eq:H-momentum-space}), where $n_q$ is the occupation number, is small compared with $\pi/2$~\cite{Imambekov2009}. We shall check this condition a posteriori.

At half filling, the gapless LL description holds for $-2<V<2$. At $V=2$, a charge density wave (CDW) gap opens up. An optimal power-law protocol for bringing the system from the gapped phase to the critical point was found in Ref.~\cite{Barankov2008} using adiabatic perturbation theory~\cite{Degrandi2010}. Here we consider the problem of transforming the system initially at the CDW phase transition critical point to a point deep within the gapless LL phase.

Let us now assume we are initially in the ground state at the critical point ($V_1=2$, $K_1=\frac{1}{2}$). We would like to find the time-dependent interaction strength $-2<V(t)<2$ for $0<t<\tau$ that yields the maximum overlap with the ground state of the Hamiltonian Eq.~(\ref{eq:Hamiltonian}) with $V=V_2$ at time $\tau$.

We proceed by expressing $\Pi_q=-i\: \partial_{\Phi_q}$ in Eq.~(\ref{eq:H-momentum-space}). The time-dependent many-body wave function of the system for a protocol $V(t)$ (and consequently $u(t)$ and $K(t)$) can then be written in the $|\{\Phi_q\}\rangle$ basis as
\begin{equation}\label{eq:Gaussian-wave-function}
\Psi(\{\Phi_q\})=\prod_{q>0}\big[  \Psi_q(\Re\:\Phi_q,t)\:\Psi_q(\Im\: \Phi_q,t)\big] 
\end{equation}
where $\Re$ and $\Im$ indicate the real and imaginary part and $\Psi_q(\phi,t)$ is the solution of the following Schr\"odinger equation
\[
\left[i \partial_t-u(t) \left(
-\frac{K(t)}{4} \:
\partial^2_{\phi}
 + \frac{1}{K(t)}\:q^2\:\phi^2
 \right)\right]\Psi_q(\phi,t)=0
\]
with appropriate initial conditions.

Up to an unimportant overall phase, the solution of the above differential equation is
\begin{equation}\label{eq:mode-wave-function}
\Psi_q(\phi,t)=\left(\frac{2\:q}{\pi}\right)^{\frac{1}{4}}\left[\Re\: z_q(t) \right]^{\frac{1}{4}}
\exp \left( -q \:z_q(t)\:\phi^2\right)
\end{equation}
where $z_q(t)$ is a complex-valued function that satisfies the following Riccati equation
\begin{equation}\label{eq:differential}
 i\: \dot{z}_q(t)=
q\:\frac{u(t)}{\alpha(t)}\:\left[z_q^2(t)-\alpha^2(t)\right]
\end{equation}
with $\alpha(t)\equiv 1/K(t)$ and the initial condition $z_q(0)=\frac{1}{K_1}$.

To perform the optimization for the many-body system, we discretize time and approximate a general $V(t)$ by a piece-wise constant function over the interval $[0\dots \tau]$. This allows us to write the final overlap, which is a functional of $V(t)$, as a multi-variable function that can be maximized numerically. An unbiased optimal protocol can be found by increasing the number of discretization points.

Note that our approach to computing the overlap could also be useful for a variety of other calculations (see~\cite{Cazalilla2006,*Iucci2009,Perfetto2010} for example) in the quench dynamics of Luttinger liquids. Let us assume a sequence $\tilde{V}_j$ with $j=1\dots N$ such that with $\Delta t={\tau}/{N}$, $V(t)=\tilde{V}_j$ for $(j-1)\Delta t<t<j\Delta t$. We then get two corresponding sequences $u_j$ and $\alpha_j$. If $z^j_q\equiv z_q(j \Delta t)$, we obtain the following recursion relation for $z^j_q$  by solving Eq.~(\ref{eq:differential}) for time-independent $u$ and $\alpha$,
\begin{equation}\label{eq:recursion}
z^{j}_q=i \: \alpha_j \: \tan\left[ q \:u_j \:\Delta t +\arctan\left( -i \frac{z^{j-1}_q}{\alpha_j }\right) \right].
\end{equation}

Our focus here is finding an optimal protocol but Eq.~(\ref{eq:recursion}) above is of interest in its own right since it gives an exact solution of the nonequilibrium wave function for any sequence of sudden quenches in the Luttinger model. Notice that knowing $z_q(t)$ for all modes determines the many-body wave function.

Recursively solving the above relation Eq.~(\ref{eq:recursion}) yields $z_q(\tau)=z^{N}_q$ for any given piece-wise constant interaction strength. Defining $\alpha_2\equiv 1/K_2$, the overlap Eq.~(\ref{eq:overlap}) can then be written as
\begin{equation}\label{eq:overlap2}
{\cal F}(\tilde{V}_1,\dots \tilde{V}_N)=\exp \left[\:\sum_{q>0}\ln \left( 4 \:\alpha_2 \:\frac{\Re\: z_q(\tau) }{|\:\alpha_2+ z_q(\tau)\:| ^2}\right)\:  \right].
\end{equation}

The overlap above is written as a multi-variable function of $\{\tilde{V}_j\}$. To find the optimal protocol, we minimize the cost function ${\cal E}(\{\tilde{V}_j\})\equiv - \ln {\cal F}(\{\tilde{V}_j\})$ with respect to the configuration $\{\tilde{V}_j\}$. This can be done by Monte-Carlo (MC) methods. We perform simulated annealing calculations with kinetic moves consisting of random small changes in randomly chosen $\tilde{V}_j$s.

We compare the results of the optimal protocol against two additional calculations. We consider the one-parameter variational protocol $V(t)=V_1+(V_2- V_1)\:(t/\tau)^r$ and calculate the final overlap for the linear ($r=1$) as well as for the the best power-law protocol ($r=r_{\rm min}$ with $\partial_r\:{\cal E}(r)=0$). The optimal protocol found by MC simulations performs significantly better than both of the above.

\begin{figure}
 \centering
 \includegraphics[width =8 cm]{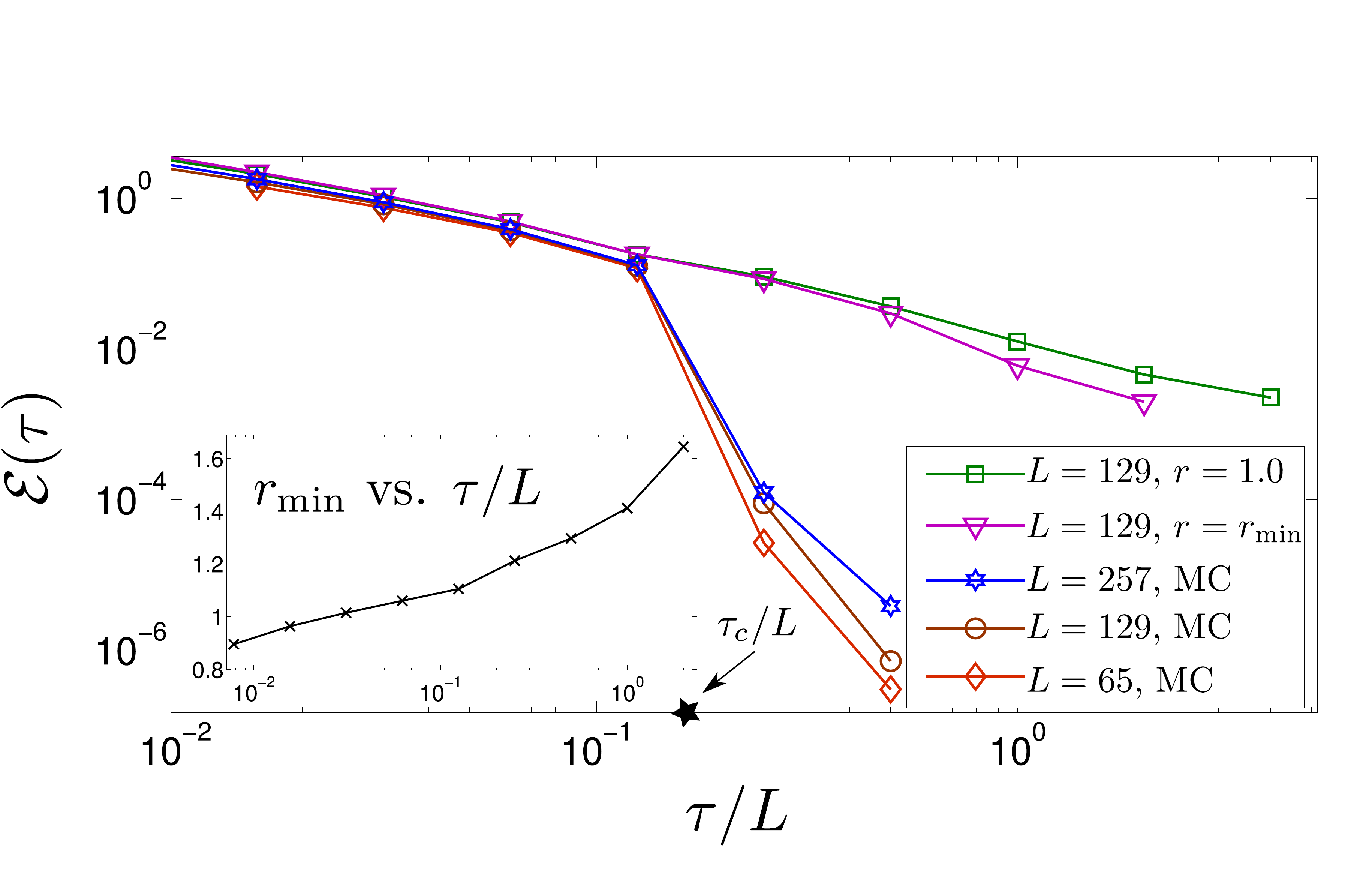}
 \caption{The final ${\cal E}$ obtained by Monte-Carlo (MC) simulations compared with the linear and best power-law protocols. The final ${\cal E}$ plunges by several orders of magnitude for $\tau/L >\tau_c/L\approx\frac{1}{8}$ (marked with a star). Note that in this regime, the data merely represents an upper bound on ${\cal E}(\tau)$ and the actual cost function could be even smaller.}
 \label{fig:optimal}
\end{figure}

In Fig.~\ref{fig:optimal}, we show the final ${\cal E}$ for the optimal protocol obtained by an unbiased MC simulation as well as for the linear and the best power-law protocols for $V_2=-1.5$ and several system sizes. When $\tau/L$ becomes larger than a critical value, the final ${\cal E}$ obtained by MC optimization exhibits a qualitative change of behavior and shoots down by several orders of magnitude. 

The value of the cost function obtained by MC simulations is generically only an upper bound on the actual minimum. For $\tau<\tau_c \propto L$, different annealing histories lead to a unique protocol suggesting that the the minimum found is likely the global minimum. For $\tau>\tau_c$ however, we never converge to a unique protocol indicating that we have only found a local minimum. In this regime, finding the global minimum becomes exceedingly difficult, particularly for larger systems due to the increased computational complexity. Nevertheless, notice that the minima we find have ${\cal E}(\tau)$ very close to zero and can prepare the new ground state almost exactly. 

Although it is not clear from the finite-size scaling of the obtained ${\cal E}(\tau)$ that there is a transition in the thermodynamic limit, the drastic change of the system behavior strongly suggests the presence of a true transition for the global minimum. Note that the local minima we find are expected to be closer to the actual global minimum for smaller systems. 

\begin{figure}
 \centering
 \includegraphics[width =8 cm]{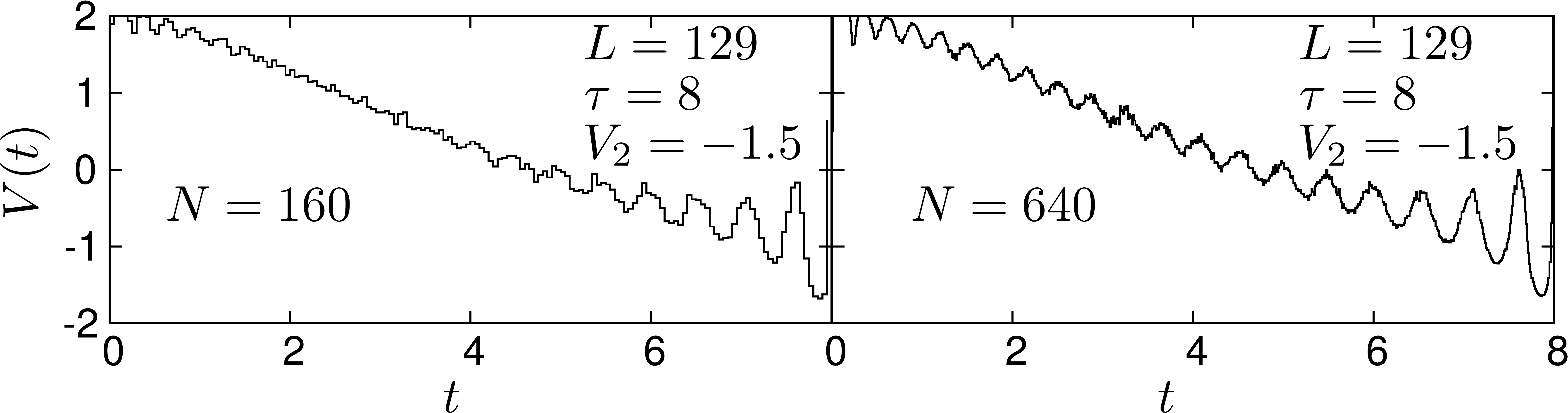}
 \caption{The interaction strength $V(t)$ as a function of time $t$ for the optimal protocol obtained by MC simulations for $L=129$, $\tau=8$ and $V_2=-1.5$. Here, the optimal protocol converges to a smooth oscillatory function as we increase the number of discretization points $N$.}
 \label{fig:protocol}
\end{figure}

\begin{figure}
 \centering
 \includegraphics[width =8 cm]{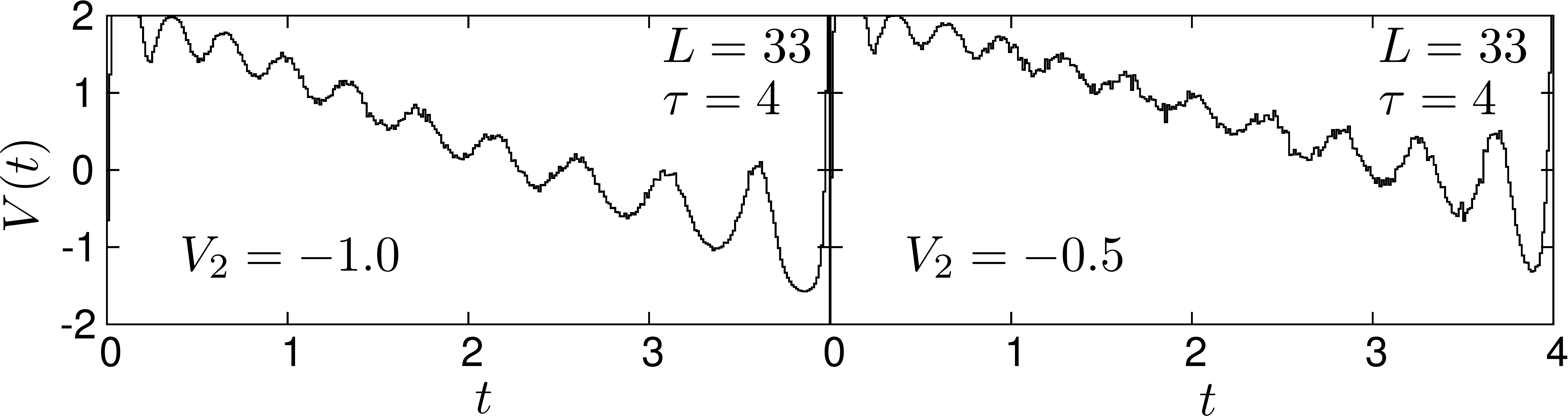}
 \caption{The optimal protocol interaction strength $V(t)$ for two different values of $V_2=-1.0$ and $V_2=-0.5$.}
 \label{fig:V_dependence}
\end{figure}

An interesting feature of the optimal protocols is that, as seen in Fig.~\ref{fig:protocol}, the interaction strength can be an oscillatory function of time. Note that increasing the number of discretization points leads to convergence to smooth but nonmonotonic protocols in the $\tau<\tau_c$ regime. For large enough systems, the period of the oscillations does not have a strong dependence on the system size $L$ or the preparation time $\tau$. With no other time scale left, we then conjecture that the oscillations must be a consequence of the short-distance length scale, i.e. the lattice spacing which is set to unity in our problem. As seen in Fig.~\ref{fig:V_dependence}, the period of the oscillations decreases as $V_2$ becomes larger. This observation is consistent with a short-distance cut-off controlling the oscillations as the velocity $u(V)$ is an increasing function of the interaction strength $V$.

We now speculate a possible scenario for the nature of the transition above. Let us consider a single mode $q$. We expect to be able to find an exact solution to $z_q(\tau)=\alpha_2$ by using a two-parameter variational protocol and solving two equations (real and imaginary part of $z_q$) and two unknowns. Generically however, this system of equations does not admit a solution. For $\tau=0$, for example, $z_q$ cannot change from $z_q(0)=\alpha_1$. By rescaling time in Eq.~(\ref{eq:differential}), we notice that the existence of a solution depends only on the quantity $\tau q$. We have checked numerically that with a two-parameter piece-wise constant protocol, we are able to obtain solutions to $z_q(\tau)=\alpha_2$ with $|\alpha_2-\alpha_1|\simeq{\cal O}(1)$ only when $q \tau\gtrsim{\cal O}(1)$.

We conjecture that even with an infinite number of variational parameters, a minimum time of ${\cal O}(1/q)$ is still required. Circumstantial support for this conjecture comes from the MC simulations with piece-wise constant protocols where, as seen in Fig.~\ref{fig:protocol} for $\tau <\tau_c$, there is convergence in the shape of the optimal protocols as we increase $N$. Notice that although we have more variation parameters $\tilde{V}_i$ for larger $N$, each of them acts for a shorter time. When $q\tau$ is large enough such that, with two variational parameters, $z_q(\tau)=\alpha_2$ has a solution, we have an infinite number of solutions for $N>2$.

For the many-mode problem, we wish to have $z_q(\tau)=\alpha_2$ for all the modes. From the discussion of the single-mode case above, we find that, for times larger than some $\tau \propto L$, an infinite number of solutions exists for the slowest mode ($q=\frac{2 \pi}{L}$) and consequently for every individual mode $q$. Since the system of equations $z_q(\tau)=\alpha_2$, $\forall q$ is under-determined when the number of variational parameters is larger than $L-1$, it seems plausible that for $\tau>\tau_c \propto L$, one can simultaneously satisfy $z_q(\tau)=\alpha_2$ for all modes. We have not, however, been able to explicitly find such exact solutions for the many-mode problem and thus cannot rule out the possibility that the transition at $\tau>\tau_c \propto L$ happens because $|z_q(\tau)-\alpha_2|$ for all modes can become infinitesimally small rather than exactly zero.

Before proceeding, let us check the applicability of the results obtained for the Luttinger model Eq.~(\ref{eq:H-momentum-space}) to the Hubbard model Eq.~(\ref{eq:Hamiltonian}). We need to calculate the total mode momentum $q n_q$. Writing the occupation number as $n_q=\langle \varepsilon_q \rangle/{2 \varepsilon^0_q}-1/2$ where $\varepsilon^0_q$ is the ground-state energy of a mode, we obtain $n_q(t)=\frac{1}{4 \Re z_q(t)}\left(K(t)\:|z_q(t)|^2+\frac{1}{K(t)} \right) -\frac{1}{2}$.
%\begin{equation}\label{eq:mode_energy}
%n_q(t)=\frac{1}{4 \Re z_q(t)}\left(K(t)\:|z_q(t)|^2+\frac{1}{K(t)} \right) -\frac{1}{2}. 
%\end{equation}
We find that when the final overlap is large, the evolution does not typically excite electrons too far away from the Fermi surface. For example, the protocol for $L=129$ and $\tau=16$ in Fig.~\ref{fig:protocol} has $\max_{q,t} \left[q n_q(t)\right]=0.302$ which is indeed in the linear regime.

From the experimental point of view, the extremely small ${\cal E}(\tau)$ obtained in the regime $\tau >\tau_c$ can be very useful for the preparation of many-body states. In practice however, there are always inaccuracies in the experimental implementation of a prescribed protocol. We check the robustness of these protocols against random perturbations $\delta V(t)$ taken from a uniform distribution $[-W/2 \dots W/2]$ for each segment of the piece-wise constant $V(t)$. We also check the effect of errors in the initial wave function by applying the protocol to the ground state for $V_1-\delta V_1$ instead of $V_1$.

\begin{figure}
 \centering
 \includegraphics[width =8 cm]{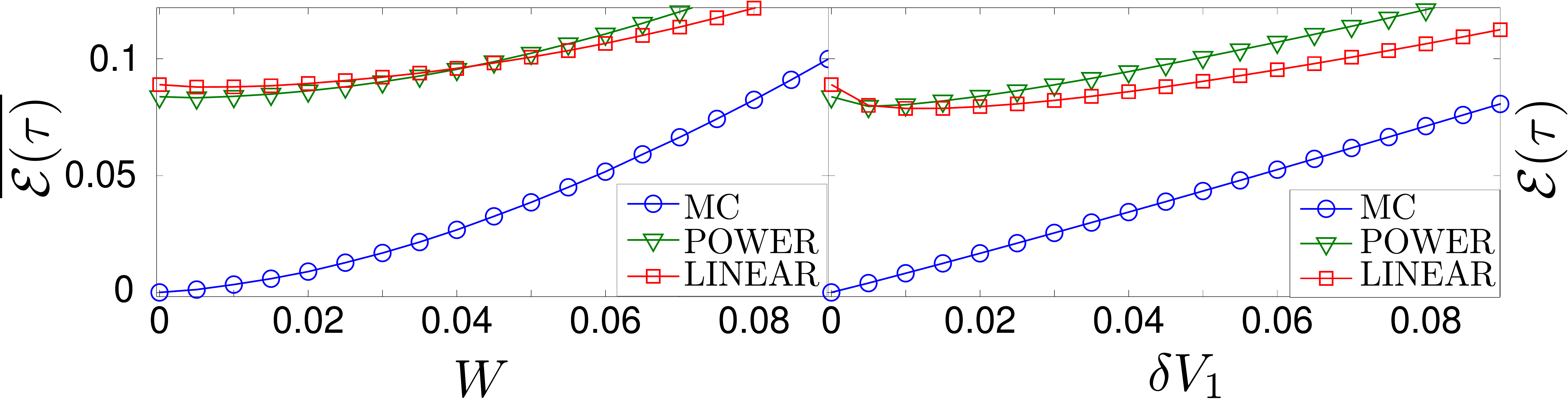}
 \caption{Left: the noise-averaged $\overline{{\cal E}(\tau)}$ as a function of $W$ for a random noise $\delta V(t)$ taken from a uniform distribution $[-W/2 \dots W/2]$. Right: final cost ${\cal E}(\tau)$ as a function of $\delta V_1$ where the initial wave function is the ground state for interaction strength $V_1-\delta V_1$. The plots are for $L=65$, $\tau=16$ and $V_2=-1.5$. }
 \label{fig:stability}
\end{figure}

 In Fig.~\ref{fig:stability}, we show ${\cal E}(\tau)$ as a function of $\delta V_1$ in addition to the noise-averaged $\overline{{\cal E}(\tau)}$ as a function of $W$ for the MC optimized, best power-law and linear protocols  and for $L=65$, $\tau=16$ and $V_2=-1.5$. As seen in the figure, even for large noise of the order a few percent of the bandwidth, the MC optimized protocols yield much smaller ${\cal E}(\tau)$ than the power-law and linear $V(t)$. These studies indicate that, practically, using the optimal protocols is advantageous even in the presence of experimental errors and inaccuracies.   

To better understand the effects of this dynamical noise on the behavior of thermally isolated Luttinger liquids, we consider a simple case. We assume we are initially in the ground state for interaction strength $V_0$ (with corresponding $\alpha_0$ and $u_0$) and $V(t)=V_0+\delta V(t)$ for a time $\tau$ where $\delta V(t)$ is a random noise of strength $W$. For $W=0$, the wave function remains in the ground state and ${\cal E}(\tau)=0$.
 For small $W$, we can linearize Eq.~(\ref{eq:differential}) and write $i \:\delta \dot{z}_q=2qu_0\:\left(\delta z_q -\delta \alpha\right)$ where $\delta z_q=z_q -\alpha_0$ and $\delta \alpha=\alpha-\alpha_0$. We then obtain
 \begin{equation}\label{eq:delta_z}
\delta z_q(\tau)=2iqu_0 \int_0^\tau dt\: e^{2iqu_0(t-\tau)}\delta \alpha(t).
\end{equation}

 Using Eq.~(\ref{eq:overlap}), we get 
$
{\cal E}(\tau)=\frac{1}{4 \alpha^2}\:\sum_{q>0} |\delta z_q(\tau)|^2+{\cal O}(\delta z^3)
$
which assuming uncorrelated noise $\langle\delta \alpha(t)\delta \alpha(t^\prime)\rangle\propto W^2 \delta(t-t^\prime)$ and using Eq.~(\ref{eq:delta_z}) above leads to 
$
\overline{{\cal E}(\tau)} \propto \tau W^2
$. Note that since $\tau_c \propto L$, strong noise limits the size of the systems which can be accurately prepared. Also notice that $|\delta z_q(\tau)|^2\propto q^2$ and in contrast to coupling to a thermal bath, the dynamical noise discussed above creates more excitations in modes with higher momentum. The effect of noise on a time-dependent optimal protocol seen in Fig.~\ref{fig:stability} is mathematically more complicated but essentially similar to this simple case. The noise-averaged $\overline{{\cal E}(\tau)}$ grows as a power law with $W$ with an exponent that depends on $V_1$ and $V_2$.

In summary, we used simulated annealing to address an outstanding problem in the nonequilibrium dynamics of interacting quantum systems, namely finding \textit{unbiased} optimal protocols for unitary preparation of strongly correlated states. We focused on transforming states in the LL phase of interacting fermions with tunable interaction strength. Unbiased optimization over all ramp shapes yields nonmonotonic optimal protocols. The behavior of the system exhibits a marked transition for $\tau>\tau_c \propto L$ where the states can be transformed with almost perfect accuracy. The optimal protocols are robust against noise, which makes them of practical experimental importance for the preparation of states in cold atomic systems. 

\begin{acknowledgments}
We are grateful to F. Burnell, C. Castelnovo, C. De Grandi, A. Hamma, P. Krapivsky, A. Polkovnikov and S. Simon for helpful discussions. Toward the completion of this work, we became aware of a recent manuscript~\cite{Eurich2010} where a similar problem is studied for small ramps in infinite dimensions with dynamical mean field theory. This work was supported by the DOE Grant
DE-FG02-06ER46316.
\end{acknowledgments}

\bibliography{optimal}{}

%Merlin.mbs v4.21 2009-07-09.
\begin{thebibliography}{10}%
\makeatletter
\providecommand \@ifxundefined [1]{%
 \ifx #1\undefined \expandafter \@firstoftwo
 \else \expandafter \@secondoftwo
\fi
}%
\providecommand \@ifnum [1]{%
 \ifnum #1\expandafter \@firstoftwo
 \else \expandafter \@secondoftwo
\fi
}%
\providecommand \enquote [1]{``#1''}%
\providecommand \bibnamefont  [1]{#1}%
\providecommand \bibfnamefont [1]{#1}%
\providecommand \citenamefont [1]{#1}%
\providecommand\href[0]{\@sanitize\@href}%
\providecommand\@href[1]{\endgroup\@@startlink{#1}\endgroup\@@href}%
\providecommand\@@href[1]{#1\@@endlink}%
\providecommand \@sanitize [0]{\begingroup\catcode`\&12\catcode`\#12\relax}%
\@ifxundefined \pdfoutput {\@firstoftwo}{%
 \@ifnum{\z@=\pdfoutput}{\@firstoftwo}{\@secondoftwo}%
}{%
 \providecommand\@@startlink[1]{\leavevmode\special{html:<a href="#1">}}%
 \providecommand\@@endlink[0]{\special{html:</a>}}%
}{%
 \providecommand\@@startlink[1]{%
  \leavevmode
  \pdfstartlink
   attr{/Border[0 0 1 ]/H/I/C[0 1 1]}%
   user{/Subtype/Link/A<</Type/Action/S/URI/URI(#1)>>}%
  \relax
 }%
 \providecommand\@@endlink[0]{\pdfendlink}%
}%
\providecommand \url  [0]{\begingroup\@sanitize \@url }%
\providecommand \@url [1]{\endgroup\@href {#1}{\urlprefix}}%
\providecommand \urlprefix [0]{URL }%
\providecommand \Eprint[0]{\href }%
\@ifxundefined \urlstyle {%
  \providecommand \doi [1]{doi:\discretionary{}{}{}#1}%
}{%
  \providecommand \doi [0]{doi:\discretionary{}{}{}\begingroup
  \urlstyle{rm}\Url }%
}%
\providecommand \doibase [0]{http://dx.doi.org/}%
\providecommand \Doi[1]{\href{\doibase#1}}%
\providecommand \bibAnnote [3]{%
  \BibitemShut{#1}%
  \begin{quotation}\noindent
    \textsc{Key:}\ #2\\\textsc{Annotation:}\ #3%
  \end{quotation}%
}%
\providecommand \bibAnnoteFile [2]{%
  \IfFileExists{#2}{\bibAnnote {#1} {#2} {\input{#2}}}{}%
}%
\providecommand \typeout [0]{\immediate \write \m@ne }%
\providecommand \selectlanguage [0]{\@gobble}%
\providecommand \bibinfo [0]{\@secondoftwo}%
\providecommand \bibfield [0]{\@secondoftwo}%
\providecommand \translation [1]{[#1]}%
\providecommand \BibitemOpen[0]{}%
\providecommand \bibitemStop [0]{}%
\providecommand \bibitemNoStop [0]{.\EOS\space}%
\providecommand \EOS [0]{\spacefactor3000\relax}%
\providecommand \BibitemShut [1]{\csname bibitem#1\endcsname}%
%</preamble>
\bibitem{Greiner2002}%
  \BibitemOpen
  \bibfield{author}{%
  \bibinfo {author} {\bibfnamefont{M.}~\bibnamefont{Greiner}}, \bibinfo
  {author} {\bibfnamefont{O.}~\bibnamefont{Mandel}}, \bibinfo {author}
  {\bibfnamefont{T.}~\bibnamefont{H\"{a}nsch}},\ and\ \bibinfo {author}
  {\bibfnamefont{I.}~\bibnamefont{Bloch}},\ }%
  \bibfield{journal}{%
  \bibinfo {journal} {Nature}\ }%
  \textbf{\bibinfo {volume} {419}},\ \bibinfo {pages} {51} (\bibinfo {year}
  {2002})%
  \bibAnnoteFile{NoStop}{Greiner2002}%
\bibitem{Kinoshita2006}%
  \BibitemOpen
  \bibfield{author}{%
  \bibinfo {author} {\bibfnamefont{T.}~\bibnamefont{Kinoshita}}, \bibinfo
  {author} {\bibfnamefont{T.}~\bibnamefont{Wenger}},\ and\ \bibinfo {author}
  {\bibfnamefont{D.~S.}\ \bibnamefont{Weiss}},\ }%
  \bibfield{journal}{%
  \bibinfo {journal} {Nature}\ }%
  \textbf{\bibinfo {volume} {440}},\ \bibinfo {pages} {900} (\bibinfo {year}
  {2006})%
  \bibAnnoteFile{NoStop}{Kinoshita2006}%
\bibitem{Sadler2006}%
  \BibitemOpen
  \bibfield{author}{%
  \bibinfo {author} {\bibfnamefont{L.~E.}\ \bibnamefont{Sadler}}, \bibinfo
  {author} {\bibfnamefont{J.~M.}\ \bibnamefont{Higbie}}, \bibinfo {author}
  {\bibfnamefont{S.~R.}\ \bibnamefont{Leslie}}, \bibinfo {author}
  {\bibfnamefont{M.}~\bibnamefont{Vengalattore}},\ and\ \bibinfo {author}
  {\bibfnamefont{D.~M.}\ \bibnamefont{Stamper-Kurn}},\ }%
  \bibfield{journal}{%
  \bibinfo {journal} {Nature}\ }%
  \textbf{\bibinfo {volume} {443}},\ \bibinfo {pages} {312} (\bibinfo {year}
  {2006})%
  \bibAnnoteFile{NoStop}{Sadler2006}%
\bibitem{Hofferberth2007}%
  \BibitemOpen
  \bibfield{author}{%
  \bibinfo {author} {\bibfnamefont{S.}~\bibnamefont{Hofferberth}}, \bibinfo
  {author} {\bibfnamefont{I.}~\bibnamefont{Lesanovsky}}, \bibinfo {author}
  {\bibfnamefont{B.}~\bibnamefont{Fischer}}, \bibinfo {author}
  {\bibfnamefont{T.}~\bibnamefont{Schumm}},\ and\ \bibinfo {author}
  {\bibfnamefont{J.}~\bibnamefont{Schmiedmayer}},\ }%
  \bibfield{journal}{%
  \bibinfo {journal} {Nature}\ }%
  \textbf{\bibinfo {volume} {449}},\ \bibinfo {pages} {324} (\bibinfo {year}
  {2007})%
  \bibAnnoteFile{NoStop}{Hofferberth2007}%
\bibitem{Bloch2008}%
  \BibitemOpen
  \bibfield{author}{%
  \bibinfo {author} {\bibfnamefont{I.}~\bibnamefont{Bloch}}, \bibinfo {author}
  {\bibfnamefont{J.}~\bibnamefont{Dalibard}},\ and\ \bibinfo {author}
  {\bibfnamefont{W.}~\bibnamefont{Zwerger}},\ }%
  \bibfield{journal}{%
  \bibinfo {journal} {Rev. Mod. Phys.}\ }%
  \textbf{\bibinfo {volume} {80}},\ \bibinfo {pages} {885} (\bibinfo {year}
  {2008})%
  \bibAnnoteFile{NoStop}{Bloch2008}%
\bibitem{Duan2003}%
  \BibitemOpen
  \bibfield{author}{%
  \bibinfo {author} {\bibfnamefont{L.-M.}\ \bibnamefont{Duan}}, \bibinfo
  {author} {\bibfnamefont{E.}~\bibnamefont{Demler}},\ and\ \bibinfo {author}
  {\bibfnamefont{M.}~\bibnamefont{Lukin}},\ }%
  \bibfield{journal}{%
  \bibinfo {journal} {Phys. Rev. Lett.}\ }%
  \textbf{\bibinfo {volume} {91}},\ \bibinfo {pages} {243202} (\bibinfo {year}
  {2003})%
  \bibAnnoteFile{NoStop}{Duan2003}%
\bibitem{Garcia-Ripoll2004}%
  \BibitemOpen
  \bibfield{author}{%
  \bibinfo {author} {\bibfnamefont{J.}~\bibnamefont{Garc\'{\i}a-Ripoll}},
  \bibinfo {author} {\bibfnamefont{M.}~\bibnamefont{Martin-Delgado}},\ and\
  \bibinfo {author} {\bibfnamefont{J.~I.}\ \bibnamefont{Cirac}},\ }%
  \bibfield{journal}{%
  \bibinfo {journal} {Phys. Rev. Lett.}\ }%
  \textbf{\bibinfo {volume} {93}},\ \bibinfo {pages} {250405} (\bibinfo {year}
  {2004})%
  \bibAnnoteFile{NoStop}{Garcia-Ripoll2004}%
\bibitem{Sorensen2010}%
  \BibitemOpen
  \bibfield{author}{%
  \bibinfo {author} {\bibfnamefont{A.~S.}\ \bibnamefont{S\o{}rensen}}, \bibinfo
  {author} {\bibfnamefont{E.}~\bibnamefont{Altman}}, \bibinfo {author}
  {\bibfnamefont{M.}~\bibnamefont{Gullans}}, \bibinfo {author}
  {\bibfnamefont{J.~V.}\ \bibnamefont{Porto}}, \bibinfo {author}
  {\bibfnamefont{M.~D.}\ \bibnamefont{Lukin}},\ and\ \bibinfo {author}
  {\bibfnamefont{E.}~\bibnamefont{Demler}},\ }%
  \bibfield{journal}{%
  \bibinfo {journal} {Phys. Rev. A}\ }%
  \textbf{\bibinfo {volume} {81}},\ \bibinfo {pages} {061603} (\bibinfo {year}
  {2010})%
  \bibAnnoteFile{NoStop}{Sorensen2010}%
\bibitem{Jaksch2005}%
  \BibitemOpen
  \bibfield{author}{%
  \bibinfo {author} {\bibfnamefont{D.}~\bibnamefont{Jaksch}}\ and\ \bibinfo
  {author} {\bibfnamefont{P.}~\bibnamefont{Zoller}},\ }%
  \bibfield{journal}{%
  \bibinfo {journal} {Ann. Phys.}\ }%
  \textbf{\bibinfo {volume} {315}},\ \bibinfo {pages} {52} (\bibinfo {year}
  {2005})%
  \bibAnnoteFile{NoStop}{Jaksch2005}%
\bibitem{Lewenstein2007}%
  \BibitemOpen
  \bibfield{author}{%
  \bibinfo {author} {\bibfnamefont{M.}~\bibnamefont{Lewenstein}}, \bibinfo
  {author} {\bibfnamefont{A.}~\bibnamefont{Sanpera}}, \bibinfo {author}
  {\bibfnamefont{V.}~\bibnamefont{Ahufinger}}, \bibinfo {author}
  {\bibfnamefont{B.}~\bibnamefont{Damski}}, \bibinfo {author}
  {\bibfnamefont{A.}~\bibnamefont{Sen}},\ and\ \bibinfo {author}
  {\bibfnamefont{U.}~\bibnamefont{Sen}},\ }%
  \bibfield{journal}{%
  \bibinfo {journal} {Adv. Phys.}\ }%
  \textbf{\bibinfo {volume} {56}},\ \bibinfo {pages} {243} (\bibinfo {year}
  {2007})%
  \bibAnnoteFile{NoStop}{Lewenstein2007}%
\bibitem{Polkovnikov2010}%
  \BibitemOpen
  \bibfield{author}{%
  \bibinfo {author} {\bibfnamefont{A.}~\bibnamefont{Polkovnikov}}, \bibinfo
  {author} {\bibfnamefont{K.}~\bibnamefont{Sengupta}},\ and\ \bibinfo {author}
  {\bibfnamefont{A.}~\bibnamefont{Silva}},\ }%
  \Eprint{http://arxiv.org/abs/1007.5331v1}{arXiv:1007.5331v1}%
  \bibAnnoteFile{NoStop}{Polkovnikov2010}%
\bibitem{Polkovnikov2008}%
  \BibitemOpen
  \bibfield{author}{%
  \bibinfo {author} {\bibfnamefont{A.}~\bibnamefont{Polkovnikov}}\ and\
  \bibinfo {author} {\bibfnamefont{V.}~\bibnamefont{Gritsev}},\ }%
  \bibfield{journal}{%
  \bibinfo {journal} {Nature Phys.}\ }%
  \textbf{\bibinfo {volume} {4}},\ \bibinfo {pages} {477} (\bibinfo {year}
  {2008})%
  \bibAnnoteFile{NoStop}{Polkovnikov2008}%
\bibitem{Eckstein2010}%
  \BibitemOpen
  \bibfield{author}{%
  \bibinfo {author} {\bibfnamefont{M.}~\bibnamefont{Eckstein}}\ and\ \bibinfo
  {author} {\bibfnamefont{M.}~\bibnamefont{Kollar}},\ }%
  \bibfield{journal}{%
  \bibinfo {journal} {New J. of Phys.}\ }%
  \textbf{\bibinfo {volume} {12}},\ \bibinfo {pages} {055012} (\bibinfo {year}
  {2010})%
  \bibAnnoteFile{NoStop}{Eckstein2010}%
\bibitem{Caneva2009}%
  \BibitemOpen
  \bibfield{author}{%
  \bibinfo {author} {\bibfnamefont{T.}~\bibnamefont{Caneva}}, \bibinfo {author}
  {\bibfnamefont{M.}~\bibnamefont{Murphy}}, \bibinfo {author}
  {\bibfnamefont{T.}~\bibnamefont{Calarco}}, \bibinfo {author}
  {\bibfnamefont{R.}~\bibnamefont{Fazio}}, \bibinfo {author}
  {\bibfnamefont{S.}~\bibnamefont{Montangero}}, \bibinfo {author}
  {\bibfnamefont{V.}~\bibnamefont{Giovannetti}},\ and\ \bibinfo {author}
  {\bibfnamefont{G.~E.}\ \bibnamefont{Santoro}},\ }%
  \bibfield{journal}{%
  \bibinfo {journal} {Phys. Rev. Lett.}\ }%
  \textbf{\bibinfo {volume} {103}},\ \bibinfo {pages} {240501} (\bibinfo {year}
  {2009})%
  \bibAnnoteFile{NoStop}{Caneva2009}%
\bibitem{Doria2010}%
  \BibitemOpen
  \bibfield{author}{%
  \bibinfo {author} {\bibfnamefont{P.}~\bibnamefont{Doria}}, \bibinfo {author}
  {\bibfnamefont{T.}~\bibnamefont{Calarco}},\ and\ \bibinfo {author}
  {\bibfnamefont{S.}~\bibnamefont{Montangero}},\ }%
  \Eprint{http://arxiv.org/abs/1003.3750v1}{arXiv:1003.3750v1}%
  \bibAnnoteFile{NoStop}{Doria2010}%
\bibitem{Rezakhani2009}%
  \BibitemOpen
  \bibfield{author}{%
  \bibinfo {author} {\bibfnamefont{A.}~\bibnamefont{Rezakhani}}, \bibinfo
  {author} {\bibfnamefont{W.-J.}\ \bibnamefont{Kuo}}, \bibinfo {author}
  {\bibfnamefont{A.}~\bibnamefont{Hamma}}, \bibinfo {author}
  {\bibfnamefont{D.}~\bibnamefont{Lidar}},\ and\ \bibinfo {author}
  {\bibfnamefont{P.}~\bibnamefont{Zanardi}},\ }%
  \bibfield{journal}{%
  \bibinfo {journal} {Phys. Rev. Lett.}\ }%
  \textbf{\bibinfo {volume} {103}},\ \bibinfo {pages} {080502} (\bibinfo {year}
  {2009})%
  \bibAnnoteFile{NoStop}{Rezakhani2009}%
\bibitem{Rahmani2010}%
  \BibitemOpen
  \bibfield{author}{%
  \bibinfo {author} {\bibfnamefont{A.}~\bibnamefont{Rahmani}}, \bibinfo
  {author} {\bibfnamefont{C.-Y.}\ \bibnamefont{Hou}}, \bibinfo {author}
  {\bibfnamefont{A.}~\bibnamefont{Feiguin}}, \bibinfo {author}
  {\bibfnamefont{C.}~\bibnamefont{Chamon}},\ and\ \bibinfo {author}
  {\bibfnamefont{I.}~\bibnamefont{Affleck}},\ }%
  \bibfield{journal}{%
  \bibinfo {journal} {Phys. Rev. Lett.}\ }%
  \textbf{\bibinfo {volume} {105}},\ \bibinfo {pages} {226803} (\bibinfo {year}
  {2010})%
  \bibAnnoteFile{NoStop}{Rahmani2010}%
\bibitem{Olshanii1998}%
  \BibitemOpen
  \bibfield{author}{%
  \bibinfo {author} {\bibfnamefont{M.}~\bibnamefont{Olshanii}},\ }%
  \bibfield{journal}{%
  \bibinfo {journal} {Phys. Rev. Lett.}\ }%
  \textbf{\bibinfo {volume} {81}},\ \bibinfo {pages} {938} (\bibinfo {year}
  {1998})%
  \bibAnnoteFile{NoStop}{Olshanii1998}%
\bibitem{Jaksch1998}%
  \BibitemOpen
  \bibfield{author}{%
  \bibinfo {author} {\bibfnamefont{D.}~\bibnamefont{Jaksch}}, \bibinfo {author}
  {\bibfnamefont{C.}~\bibnamefont{Bruder}}, \bibinfo {author}
  {\bibfnamefont{J.~I.}\ \bibnamefont{Cirac}}, \bibinfo {author}
  {\bibfnamefont{C.~W.}\ \bibnamefont{Gardiner}},\ and\ \bibinfo {author}
  {\bibfnamefont{P.}~\bibnamefont{Zoller}},\ }%
  \bibfield{journal}{%
  \bibinfo {journal} {Phys. Rev. Lett.}\ }%
  \textbf{\bibinfo {volume} {81}},\ \bibinfo {pages} {3108} (\bibinfo {year}
  {1998})%
  \bibAnnoteFile{NoStop}{Jaksch1998}%
\bibitem{Giorgini2008}%
  \BibitemOpen
  \bibfield{author}{%
  \bibinfo {author} {\bibfnamefont{S.}~\bibnamefont{Giorgini}}, \bibinfo
  {author} {\bibfnamefont{L.}~\bibnamefont{Pitaevskii}},\ and\ \bibinfo
  {author} {\bibfnamefont{S.}~\bibnamefont{Stringari}},\ }%
  \bibfield{journal}{%
  \bibinfo {journal} {Rev. Mod. Phys.}\ }%
  \textbf{\bibinfo {volume} {80}},\ \bibinfo {pages} {1215} (\bibinfo {year}
  {2008})%
  \bibAnnoteFile{NoStop}{Giorgini2008}%
\bibitem{Jordens2008}%
  \BibitemOpen
  \bibfield{author}{%
  \bibinfo {author} {\bibfnamefont{R.}~\bibnamefont{J\"{o}rdens}}, \bibinfo
  {author} {\bibfnamefont{N.}~\bibnamefont{Strohmaier}}, \bibinfo {author}
  {\bibfnamefont{K.}~\bibnamefont{G\"{u}nter}}, \bibinfo {author}
  {\bibfnamefont{H.}~\bibnamefont{Moritz}},\ and\ \bibinfo {author}
  {\bibfnamefont{T.}~\bibnamefont{Esslinger}},\ }%
  \bibfield{journal}{%
  \bibinfo {journal} {Nature}\ }%
  \textbf{\bibinfo {volume} {455}},\ \bibinfo {pages} {204} (\bibinfo {year}
  {2008})%
  \bibAnnoteFile{NoStop}{Jordens2008}%
\bibitem{Imambekov2009}%
  \BibitemOpen
  \bibfield{author}{%
  \bibinfo {author} {\bibfnamefont{A.}~\bibnamefont{Imambekov}}\ and\ \bibinfo
  {author} {\bibfnamefont{L.~I.}\ \bibnamefont{Glazman}},\ }%
  \bibfield{journal}{%
  \bibinfo {journal} {Science}\ }%
  \textbf{\bibinfo {volume} {323}},\ \bibinfo {pages} {228} (\bibinfo {year}
  {2009})%
  \bibAnnoteFile{NoStop}{Imambekov2009}%
\bibitem{Barankov2008}%
  \BibitemOpen
  \bibfield{author}{%
  \bibinfo {author} {\bibfnamefont{R.}~\bibnamefont{Barankov}}\ and\ \bibinfo
  {author} {\bibfnamefont{A.}~\bibnamefont{Polkovnikov}},\ }%
  \bibfield{journal}{%
  \bibinfo {journal} {Phys. Rev. Lett.}\ }%
  \textbf{\bibinfo {volume} {101}},\ \bibinfo {pages} {076801} (\bibinfo {year}
  {2008})%
  \bibAnnoteFile{NoStop}{Barankov2008}%
\bibitem{Degrandi2010}%
  \BibitemOpen
  \bibfield{author}{%
  \bibinfo {author} {\bibfnamefont{C.}~\bibnamefont{De~Grandi}}\ and\ \bibinfo
  {author} {\bibfnamefont{A.}~\bibnamefont{Polkovnikov}},\ }%
  \bibfield{journal}{%
  \bibinfo {journal} {Lect. Notes Phys.}\ }%
  \textbf{\bibinfo {volume} {802}},\ \bibinfo {pages} {75} (\bibinfo {year}
  {2010})%
  \bibAnnoteFile{NoStop}{Degrandi2010}%
\bibitem{Cazalilla2006}%
  \BibitemOpen
  \bibfield{author}{%
  \bibinfo {author} {\bibfnamefont{M.}~\bibnamefont{Cazalilla}},\ }%
  \bibfield{journal}{%
  \bibinfo {journal} {Phys. Rev. Lett.}\ }%
  \textbf{\bibinfo {volume} {97}},\ \bibinfo {pages} {156403} (\bibinfo {year}
  {2006})%
  \bibAnnoteFile{NoStop}{Cazalilla2006}%
\bibitem{Iucci2009}%
  \BibitemOpen
  \bibfield{author}{%
  \bibinfo {author} {\bibfnamefont{A.}~\bibnamefont{Iucci}}\ and\ \bibinfo
  {author} {\bibfnamefont{M.~A.}\ \bibnamefont{Cazalilla}},\ }%
  \bibfield{journal}{%
  \bibinfo {journal} {Phys. Rev. A}\ }%
  \textbf{\bibinfo {volume} {80}},\ \bibinfo {pages} {063619} (\bibinfo {year}
  {2009})%
  \bibAnnoteFile{NoStop}{Iucci2009}%
\bibitem{Perfetto2010}%
  \BibitemOpen
  \bibfield{author}{%
  \bibinfo {author} {\bibfnamefont{E.}~\bibnamefont{Perfetto}}, \bibinfo
  {author} {\bibfnamefont{G.}~\bibnamefont{Stefanucci}},\ and\ \bibinfo
  {author} {\bibfnamefont{M.}~\bibnamefont{Cini}},\ }%
  \bibfield{journal}{%
  \bibinfo {journal} {Phys. Rev. Lett.}\ }%
  \textbf{\bibinfo {volume} {105}},\ \bibinfo {pages} {156802} (\bibinfo {year}
  {2010})%
  \bibAnnoteFile{NoStop}{Perfetto2010}%
\bibitem{Eurich2010}%
  \BibitemOpen
  \bibfield{author}{%
  \bibinfo {author} {\bibfnamefont{N.}~\bibnamefont{Eurich}}, \bibinfo {author}
  {\bibfnamefont{M.}~\bibnamefont{Eckstein}},\ and\ \bibinfo {author}
  {\bibfnamefont{P.}~\bibnamefont{Werner}},\ }%
  \Eprint{http://arxiv.org/abs/1010.2853v1}{arXiv:1010.2853v1}%
  \bibAnnoteFile{NoStop}{Eurich2010}%
\end{thebibliography}%

\end{document}

\maketitle
\begin{abstract}
The trajectories of $z_q(t)$ on the complex $z$ plane are discussed for the evolution with optimal protocols.
\end{abstract}

\begin{figure}
 \centering
 \includegraphics[width =12 cm]{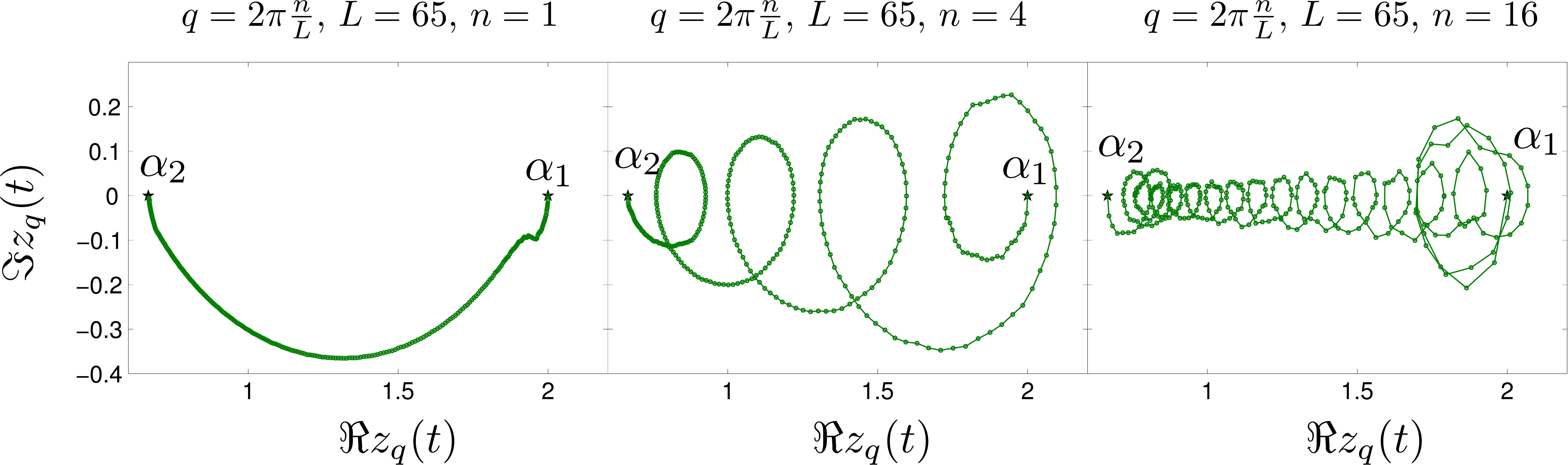}
 \caption{The values of $z_q(t)$ for three different modes during the evolution with the MC optimized protocol with $L=65$, $\tau=16$ and $V_2=-1.5$. As the momentum $q$ increases the modes wind around in the complex plane to reach the point $\alpha_2$. }
 \label{fig:winding}
\end{figure}

Some intuition may be gained regarding the appearance of oscillations in optimal protocols discussed in the main text by considering the trajectories of $z_q(t)$ on the complex $z$ plane. The evolution of the system is governed by a set of nonlinear differential equations (Eq.~(6) of the main text) for many modes. We would like controls $\alpha(t)$ and $u(t)$ such that for all modes $q$, the complex function $z_q(t)$ moves from the initial value $z_q(0)=\alpha_1$ to a desired final value $\alpha_2$ in a time $\tau$. Now the quantity controlling this evolution for each mode is $q \tau$ as discussed in the main text. It is evident from Eq.~(6) of the main text that $z_q$ moves faster for modes with higher momenta. To get to the same point $\alpha_2$, we find that $z_q$ actually winds around in the complex plane as seen in Fig.~\ref{fig:winding}. The appearance of these windings, i.e oscillations in the imaginary part of $z_q$, is rather intuitive. One may guess that an oscillatory control is required to generate the winding of the complex $z_q(t)$.
